\documentclass{sig-alternate-05-2015}
\usepackage{letltxmacro}
\usepackage{cite}
\usepackage{lipsum}
\usepackage{multicol}
\usepackage{graphicx}

\LetLtxMacro{\OldIncludegraphics}{\includegraphics}

\def\sharedaffiliation{%
\end{tabular}
\begin{tabular}{cc}}

\begin{document}


\permission{Permission to make digital or hard copies of part or all of this work for personal or classroom use is granted without fee provided that copies are not made or distributed for profit or commercial advantage and that copies bear this notice and the full citation on the first page. Copyrights for third-party components of this work must be honored. For all other uses, contact the owner/author(s). \\ 
\\
{\small \textit{ Neu-IR '16 SIGIR Workshop on Neural Information Retrieval, July 21, 2016, Pisa, Italy. } }}
\conferenceinfo{Neu-IR '16 SIGIR Workshop on Neural Information Retrieval }{July 21, 2016, Pisa, Italy.}
\copyrightetc{\copyright 2016 Copyright held by the owner/author(s).}

\title{Deep Feature Fusion Network for Answer Quality Prediction in Community Question Answering}
\numberofauthors{4}
    \author{
      \alignauthor Sai Praneeth Suggu$^{\star}$      
      \alignauthor Kushwanth N. Goutham$^{\star}$
\and
      \alignauthor Manoj K. Chinnakotla$^{\dagger}$
      \alignauthor Manish Shrivastava$^{\star}$
      \sharedaffiliation
      \affaddr{}\\
      \affaddr{$^{\star}$IIIT Hyderabad, India}\\
      \affaddr{\texttt{\{suggusai.praneeth,kushwanth.naga\}@research.iiit.ac.in}}\\
      \affaddr{\texttt{m.shrivastava@iiit.ac.in}}\\\\
      \affaddr{$^{\dagger}$Microsoft, India}\\
      \affaddr{\texttt{manojc@microsoft.com}}
      }
%
 
\maketitle
\begin{abstract}
Community Question Answering (cQA) forums have become a popular medium for soliciting direct answers to specific questions of users from experts or other experienced users on a given topic. However, for a given question, users sometimes have to sift through a large number of low-quality or irrelevant answers to find out the answer which satisfies their information need. To alleviate this, the problem of Answer Quality Prediction (AQP) aims to predict the quality of an answer posted in response to a forum question. Current AQP systems either learn models using - a) various hand-crafted features (HCF) or b) use deep learning (DL) techniques which automatically learn the required feature representations.

In this paper, we propose a novel approach for AQP known as - \emph{``Deep Feature Fusion Network (DFFN)''} which leverages the advantages of both hand-crafted features and deep learning based systems. Given a question-answer pair along with its metadata, DFFN independently -  a) learns deep features using a Convolutional Neural Network (CNN) and b) computes hand-crafted features using various external resources and then combines them using a deep neural network trained to predict the final answer quality. DFFN achieves \emph{state-of-the-art performance} on the standard SemEval-2015 and SemEval-2016 benchmark datasets and outperforms baseline approaches which individually employ either HCF or DL based techniques alone.
\end{abstract}


\section{Introduction}
Community Question Answering (cQA) forums (such as \emph{Yahoo! Answers}, \emph{Stack Overflow}, \emph{etc.}) have become a popular medium for many internet users to get precise answers or opinions to their questions from experts or other experienced users in the topic. Such forums are usually open, allowing any user to respond to a given question. As a result, for a given question posed by the user, the quality of response often varies a lot ranging from highly precise and detailed answers from authentic users to highly imprecise or non-comprehensible one-word, single line answers or answers which are completely unrelated to the topic posted by spammy and other non-serious users. This severely hampers the effectiveness of the cQA forums as users will have to sift through a large number of irrelevant posts to find the answers satisfying their information needs. To alleviate this problem, cQA forums often include feedback mechanisms such as votes, ratings \emph{etc.} for rating the quality of answers and users which could also be used as signals for ranking the answers given a question. However, such popularity based signals (votes, ratings) are often prone to spam due to users who may artificially inflate their ratings, votes with the help of other users whom they know. To overcome the above problems, recent approaches \cite{jaist,hitsz-icrc,qcri,ecnu,wang:2009,icrc-hit,DBLP:journals/corr/HsuZG16,rankshorttext,DBLP:journals/corr/YuHBP14} have focused on automatically ranking answers for a given question based on their quality.

The problem of answer quality prediction is defined as follows: Given a question $Q$ and its set of community answers $C=\{A_1,A_2,\ldots,A_n\}$, rate the answers corresponding to their quality. The cQA tasks of SemEval-2015 (Task A) \cite{semeval15task3} and SemEval-2016 (Task A) \cite{semeval16task3} provide a universal benchmark for evaluating research on this problem. In SemEval-2015, the answers are to be rated as \textit{\{good,potentially useful or bad\}} and in SemEval-2016, the answers are to be rated as either \textit{\{good or bad}\}.

Recent approaches for answer quality prediction can be categorized into - a) Hand-crafted Feature (HCF) based approaches \cite{jaist,hitsz-icrc,qcri,ecnu,wang:2009} or b) Deep Learning (DL) based approaches  \cite{icrc-hit,DBLP:journals/corr/HsuZG16,rankshorttext,DBLP:journals/corr/YuHBP14}. HCF based approaches mainly rely on capturing various semantic and syntactic similarities between the question and answer, behavior of users using feature engineering. For computing these similarities, recent approaches have leveraged external knowledge resources such as WordNet and other text corpora. DL based approaches, on the other hand, automatically learn the feature representations while learning the target quality scoring function. As a result, they are language-agnostic and don't require feature engineering or any external resources except for a large training corpus.
\begin{figure*}[!tp]
  \includegraphics[width=\textwidth,height=9cm]{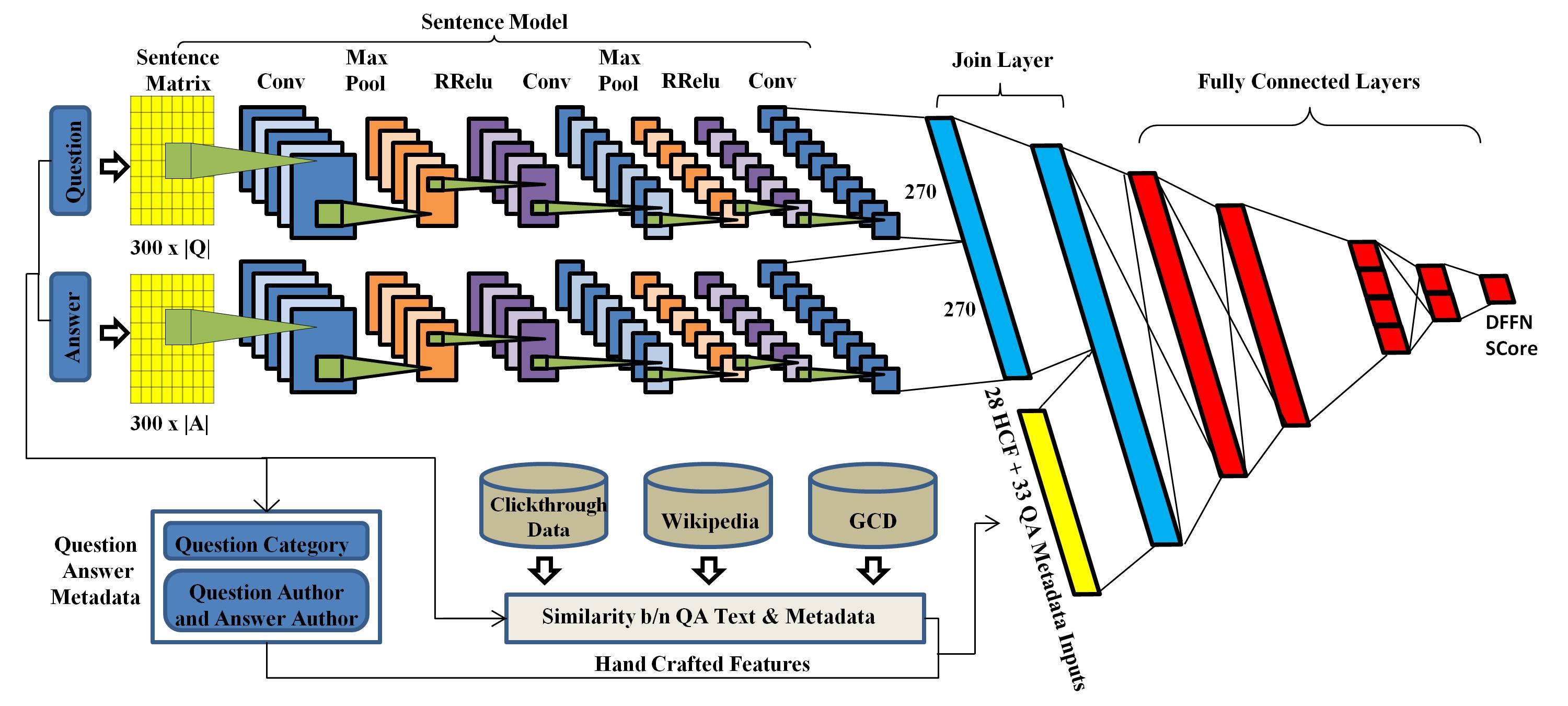}
  \label{fig:sysarch}
  \caption{System Architecture of Deep Feature Fusion Network (DFFN)}
\end{figure*}

In this paper, we propose ``Deep Feature Fusion Network (DFFN)'' - a novel approach which combines HCF into a Convolutional Neural Network (CNN) model for improving answer quality prediction. DFFN leverages the advantage of both HCF and DL based approaches \emph{i.e.} ability to - a) encode similarities between question-answer pair using external knowledge resources such as Wikipedia, Google Cross-Lingual Dictionary (GCD), Clickthrough data and b) automatically learn features and the target function. Given a question, answer pair along with its metadata, DFFN independently learns deep features from CNN, computes hand-crafted features using various external resources and then combines the deep features and hand-crafted features using a deep neural network trained to predict the quality rating of the answer. DFFN achieves state-of-the-art performance on the standard SemEval-2015 and SemEval-2016 benchmark datasets and shows better performance than baseline approaches which individually employ either HCF or DL based techniques. In this context, the following are our main contributions:
\begin{itemize}
\item We propose a novel approach to combine hand-crafted features into a CNN for the answer quality prediction task
\item We achieve state-of-the-art performance on SemEval 2015 and SemEval 2016 cQA answer quality prediction tasks
\end{itemize}
The rest of the paper is organized as follows: Section \ref{relwork} discusses the related work in this area. Section \ref{dffn} presents our contribution DFFN in detail. Section \ref{experiments} discusses our experimental set-up. Section \ref{results} presents our results and finally Section \ref{conclusion} concludes the paper.

\section{Related Work}\label{relwork}
AQP in cQA forums has been researched a lot in the IR community. Jeon \emph{et al.} \cite{sigir06} employ non-textual features such as clicks, print counts, copy counts \emph{etc.} to predict the quality of a answer in a cQA forum. Liu \emph{et al.} \cite{sigir08} investigate a slightly related problem i.e. predicting whether an asker would be satisfied with the answers provided so far to the given question. Burel \emph{et al.} \cite{Burel2012} have used a combination of content, user and thread related features for predicting answer quality. Dalip \emph{et al.} \cite{Dalip2013} propose a learning to rank approach for AQP using eight different groups of features. Li \emph{et al.} \cite{Li2015} studied the various factors such as shorter length, authors reputation which lead to a high answer quality rating as rated by peers.

More recently, Tran \emph{et al.} \cite{jaist} made use of topic models, word vectors and other hand crafted rules to train a SVM classifier for AQP.  Hou  \emph{et al.} \cite{hitsz-icrc} made use of statistics like avg. word length of a sentence (question or answer), sentence length with other hand-crafted features to train an ensemble of classifiers for AQP. Wang \emph{et.al} \cite{wang:2009} use Bayesian logistic regression and  link prediction models for AQP. Yu  \emph{et al.} \cite{DBLP:journals/corr/YuHBP14} used CNN to learn a distributional sentence model for AQP from bag of words and bigram based word representations. Nicosia \emph{et.al} \cite{qcri} have used lexical similarity between word n-grams, tree kernels, word-embeddings and other hand crafted features for AQP. Hsu \emph{et al.} \cite{DBLP:journals/corr/HsuZG16} used a LSTM encoder with neural attention mechanism to automate feature engineering process.  Severyn \emph{et al.} \cite{rankshorttext} used a CNN to automatically learn features for matching short text pairs. Zhou \emph{et al.} \cite{icrc-hit} used a 2-dimensional CNN to represent a question-answer pair and ranked the representations using a Recurrent Neural Network. 

Our work resembles the work by Wu \emph{et al.} \cite{vision_convl} who employ the idea of combining hand-crafted features and deep features for person re-identification task in computer vision. However, in our case, the idea of using hand-crafted features is driven by the availability of large similarity resources such as Wikipedia, Google Cross-Lingual Dictionary and Clickthrough data which could be leveraged to infer richer syntactic and semantic similarities between textual elements.

\section{{Deep Feature Fusion Network\\ (DFFN)}}\label{dffn}
Figure \ref{fig:sysarch} shows the architecture of our DFFN model. The input to DFFN is the question, answer and metadata (question category, question author, answer author \emph{etc.}) and the output is a relevance score depicting the quality of the answer. DFFN is a two-staged deep Neural Network (NN) model. In the first stage, DFFN has two parallel CNN based sentence models for the question and answer which are used to learn their feature representations (lets call it CNNFR). In parallel, DFFN also generates hand-crafted feature representations from the question, answer and metadata information using the Wikipedia and other similarity models (lets call it HCFR). In the second stage, CNNFR and HCFR along with Metadata (question author,answer author and question category) are joined and passed through one more deep Neural Network (NN) which predicts the score representing answer quality. We will now discuss each stage of DFNN in detail.

\subsection{Sentence Model}
The sentence model projects a sentence (question/answer) into the semantic space and learns a good intermediate representation of the given question/answer. The sentence model is a deep convolutional neural network (CNN). Our CNN mainly consists of sentence matrix and multiple convolutional, pooling and non-linearity layers as in Figure \ref{fig:sysarch}. 

The input to the sentence matrix $S$ is a vector of words from the sentence (question/answer) $s$ = $\{w_1,w_2,....w_{\left|s\right|}\}$. We build the sentence matrix by mapping each word $w_i$ in the question/answer to its corresponding word embedding in $d$ dimensions. We use GLoVE \cite{pennington2014glove} based embeddings of 300 dimensions to map the words in the question and answer. We limit the size of the sentence upto certain threshold. We ignore the words in the sentence after a certain threshold if the length of the sentence is greater than the threshold and pad zeros upto the threshold if the length of the sentence is less than the threshold. The sentence matrix is then convolved through multiple convolution, pooling and non-linearity layers to get the feature representations of the question/answer. We perform convolution in 2 dimensions i.e. horizontally and vertically. We use  max-pooling for the pooling layer and Randomized Leaky Rectified Linear Unit (RReLU) \cite{rrelu}, a randomized version of leakyReLU \cite{rrelu}, as the non-linearity layer. RReLU for a value $x$ is computed as follows in training phase:

\[
    f(x)= 
\begin{cases}
    x,&\text{if } x \geq 0\\
    ax,       & if x < 0
\end{cases}
\]
where 
\[
a \sim U(l, u), l <  u \ and \ l,u \in [0,1)
\]
\emph{i.e.} $a$ is a random number drawn from a uniform random distribution $U(l, u)$. In testing phase it is computed as:

\[
f(x)=\frac{x}{\frac{l+u}{2}}
\]

Using sentence models, we get the individual feature representations (270 dimensions) of the question and answer which are then concatenated to produce a combined feature representation (540 dimensions). These are concatenated with hand crafted features and metadata and are given as input to the second stage Neural Network. We describe in detail regarding them in the following subsections.

\subsection{Hand Crafted Features (HCF)}
The question and answer text usually consist of several Named Entities (NEs) and concepts along with their various variants. For example, the cricketer \emph{Sachin Tendulkar} could be referred to as \emph{Sachin}, \emph{Tendulkar}, \emph{The Little Master} \emph{etc.} Such variants are hard to capture using CNN based features alone. Hence, we make use of resources such as Wikipedia, Google Cross-Lingual Dictionary (GCD), Named Entity Recognizers (NER) and Clickthorugh data to come up with hand-crafted features which can capture such rich similarities. We also observe that user behaviour and specific patterns on metadata and question-answer text are useful. We use these features to compute individual similarity scores between question and answer and combine these scores as Hand Crafted Features to give them as input to the second stage NN. We describe the details of the features below:

\subsubsection{Wikipedia Based Features}
In this section, we describe the similarity features which are computed based on using Wikipedia as a resource.

\textbf{TagMe Similarity:} We extract TagMe concepts of the question and answer by mapping them to their corresponding Wikipedia page titles using TagMe \cite{tagme}. TagMe identifies meaningful substrings in an unstructured text and links them to their relevant wikipedia pages. We compute the similarity between two TagMe concepts using wikipedia Miner \cite{wikiminer}. Wikipedia Miner computes similarity between two wikipedia pages based on the number of common inlinks and outlinks between them. 

Similarity between question and answer represented by TagMe concepts using Wikipedia Miner is computed as the mean average of the similarity between pairs of TagMe concepts (one each from the question and the answer) as in Equation \ref{eq:1}
\begin{equation} \label{eq:1}
\frac{\sum_{i=1}^{n} \sum_{j=1}^{m} sim(c_i,c_j)}{nm}
\end{equation}
where $n,m$ are the number of TagMe concepts in the question and answer respectively, $c_i,c_j$ are the $i^{th}$ and $j^{th}$ TagMe concepts in the question and answer respectively, $ sim(c_i,c_j)$ is the similarity between $c_i$ and $c_j$ calculated using Wikipedia Miner.

\textbf{GCD Similarity:} Google Cross-Lingual Dictionary (GCD)\cite{gcd} is a string to concept mapping on the vast link structure of the web, created using anchor text from various pages across the web. A concept is an individual Wikipedia document. The text strings constitute the anchor texts that refer to these concepts. Thus, each anchor text string represents a concept.

We extract common and proper nouns from the question and answer using  Stanford CoreNLP POS Tagger \cite{corenlppos} and query them individually on GCD anchor texts to get top ten unique concepts related to question and answer. We calculate the similarity between two GCD concepts using Wiki Miner. The similarity between question and answer represented by GCD concepts is calculated as in Equation \ref{eq:1} where we use GCD concepts instead of TagMe concepts.

\smallskip

\textbf{Named Entities Similarity:} We extract Named Entities from the question and answer, using Stanford CoreNLP POS Tagger. We compute the similarity between two Named Entities using a GCD based similarity feature.

The GCD based similarity between two Named Entities is computed as the ratio of number of wikipedia documents in which these two terms co-occur in the top k retrieved documents when queried on GCD. For our experiments, we set k to 100. We calculated the co-occurrence of two terms in a document by checking if both the terms match with any of the words in the document with a string edit distance greater than a threshold.

Similarity between question and answer represented by Named Entities is calculated as in Equation \ref{eq:1} where we use Named Entities instead of TagMe concepts and GCD based similarity feature instead of Wikipedia Miner to calculate the similarity between two Named Entities.

\begin{table*}[!t]
\centering
\begin{tabular}{l l l| l l l l }
\hline
\multicolumn{2}{c}{\textbf{SemEval 15}} & \multicolumn{5}{c}{\textbf{SemEval 16}}                \\ \hline
\textbf{Model} & \textbf{F1} &  \textbf{Acc.} & \textbf{Model} & \textbf{MAP} & \textbf{F1} & \textbf{Acc.}\\
\hline
DFFN & \textbf{61.22*} & \textbf{75.24*} & DFFN & \textbf{82.34*} & \textbf{66.22*} & \textbf{76.67} \\
JAIST & 57.29 & 72.67 & Kelp & 79.19 &  64.36 & 75.11 \\
HITSZ-ICRC & 56.44 & 69.43 & ConvKN & 78.71 & 63.55 & 74.95 \\
DFFN \emph{w/o} HCF & 56.04  & 69.73 & DFFN \emph{w/o} HCF & 74.36 &  60.22 & 72.88 \\
DFFN \emph{w/o} CNN & 51.65 & 67.12 & DFFN \emph{w/o} CNN & 70.21 & 56.77 & 68.65 \\
ICRC-HIT & 53.82 & 73.18 & & & & \\
\hline
\end{tabular}
\caption{Overall Results of DFFN on SemEval 2015 and SemEval 2016 datasets. Results marked with a $^*$ were found to be statistically significant with respect to the nearest baseline at 95\% confidence level ($\alpha=0.05$) when tested using paired two-tailed t-test.}
\label{overallres}
\end{table*}


\subsubsection{Sentence-Vector Features}
\smallskip

\textbf{Paragraph2vec Similarity:} Paragraph2Vec\cite{para2vec} allows to model vectors for text of any arbitrary length. It learns continuous distributed vector representations for pieces of texts. We train the para2vec model on the training data of the particular tasks only (SemEval'15 and SemEval'16) by treating each question-answer pair as a single document. We train only on the good question-answer pairs from the training data.  A good question-answer pair is a pair in which answer is rated as a ``good" answer for that question. 

We map the question and answer to vectors using para2vec and compute the similarity between the question and answer as the cosine similarity between their para2vec vectors.
\smallskip

\textbf{Sent2Vec Similarity:} Sent2Vec maps a pair of short texts to a pair of feature vectors in a continuous, low-dimensional space. Sent2Vec performs the mapping using the Deep Structured Semantic Model (DSSM) built using Clickthrough data \cite{dssm}, or the DSSM with convolutional-pooling structure (CDSSM) \cite{cdssm1,cdssm2}. 

We map the question and answer to vectors using both DSSM and CDSSM. We compute the Sent2Vec DSSM similarity between the question and answer as the cosine similarity between the vectors of question and answer obtained by using Sent2Vec performing the mapping of vectors using DSSM. Similarly by using CDSSM instead of DSSM we also compute the Sent2Vec CDSSM similarity between the question and answer. 

\subsubsection{Metadata Based Features}
\smallskip

\textbf{Author Reputation Score:} We observed that the reputation of an answer author, within a forum plays a key role in determining the quality of answer. We capture this through a author reputation feature. We have two reputation features namely Good Reputation and Bad Reputation. Good reputation of an author is computed as the ratio of the number of good answers given by that author to the maximum number of good answers given by any individual author in the entire forum. Similarly, by using the number of bad answers instead of good answers, we also compute a score for the bad reputation of an author.
\smallskip

\textbf{Is Answer Seeker?:} We have a boolean feature to represent whether the answer (comment) is written by the person who has asked the question.
\smallskip

\textbf{Question Authors' Response Pattern:} We compute features based on whether the question author has commented before or after the present answer and if that comment by the question author is a question. Usually, the question author posts comments/questions below an answer if one is not satisfied with the current answer. These features capture the behavior.

\textbf{Miscellaneous:} Besides, we extract and add features related to - a) statistics of each question category (number of good, potential and bad answers in that category ) b) position of the answer. 
c) presence of  URL, e-mail in the answer d) presence of question marks, exclamation marks in the answer e) boolean features for the presence of various emoticons such as happy ( eg: ``:)", ``:D" ), sad ( eg: ``:(" , ``:'(" ) in the answer.

We obtain the similarity scores and together call them as Hand-crafted features (28 dimensions). We join them as a vector and give them as input to second stage NN along with Metadata as described in the following subsections.

\subsection{Metadata}
We observe that category of the question plays an important role in computing the answer quality score as it is easy to write good answers for some categories and difficult for some other. We also include author information of the question and answer. We encode the question category, question author and answer author using a logarithmic function and give them as input to second stage NN.

\subsection{Second Stage Neural Network}
As discussed, the vector representations from the sentence models (540 dimensions), the feature representations from HCF (28 dimensions) and direct inputs from Metadata (33 dimensions) are combined to get a single feature vector of 601 dimensions. This vector is given as input to the second stage NN consisting of fully connected layers. These layers model various interactions between the features present in the vector and finally output a score predicting the answer quality.


\subsection{Training}
The parameters of the network are learnt with an objective to maximize the accuracy of prediction given the target categories. For example, in SemEval-2015, the target categories were $\{good,potentially$  $useful,bad\}$ and $\{good,bad\}$ in SemEval-2016 . For training, we used the training data provided in the SemEval 2015 \cite{semeval15task3} and 2016 \cite{semeval16task3} tasks which consists of question, answer, metadata along with their ideal quality rating. We tuned the DFFN parameters on the corresponding development sets of SemEval 2015 and 2016. We used Adagrad\cite{adagrad} to speed up the convergence rate of stochastic gradient descent (SGD).

Given an input ${(p, t)}$ where $p$ is the predicted answer quality score by DFFN and $t$ is the true label depicting answer quality, we used SmoothL1 as the loss criterion which is computed as:
\[
    loss(p, t)= \frac{1}{n} \times 
\begin{cases}
    \ 0.5\times(p - t)^2,&\text{if } |p-t|<1\\
     |p - t| - 0.5,       & \text{if } |p-t| \geq 1
\end{cases}
\]
$t$ is 1 for good question-answer pair (answer labeled as good for that question) and -1 for bad question answer pair (answer labeled as bad for that question). The model is trained by minimizing the loss function in a batch of size $n$.

\section{Experimental Setup}\label{experiments}
We use the SemEval 2016 \cite{semeval16task3} and SemEval 2015 \cite{semeval15task3} datasets for our experiments as it exactly matches our problem description. SemEval 2016 consists of 36198 training question-answer (QA) pairs and 2440 for dev and 3270 for testing purposes. SemEval 2015 consists of 16541 training QA pairs and 1645 dev and 1976 for testing. 

To evaluate the performance, we use standard evaluation metrics - Mean Average Precision (MAP), F1 score and Accuracy. We compare our approach with the top two best performing systems from SemEval 2015 - JAIST\cite{jaist} and HITSZ-ICRC \cite{hitsz-icrc}. JAIST and HITSZ-ICRC use hand-crafted feature based models. We also compare with ICRC-HIT\cite{icrc-hit} as it uses a purely deep learning based model. Similarly, for SemEval 2016, we compare with their corresponding top two best performing systems - Kelp\cite{semeval16task3} and ConvKN \cite{semeval16task3}. We do not know their system descriptions and exact algorithms since the conference proceedings are not yet out.

\section{Results and Discussion}\label{results}
Table \ref{overallres} shows the overall results of DFFN on SemEval 2015 and SemEval 2016 datasets. DFFN performs better than the top systems across all the metrics. The improvement is higher in SemEval 2015 although the task is more harder due to lesser training data and more granularity in target labels to be predicted. We also observe that DFFN performs better than a single CNN alone (DFFN \emph{w/o} HCF) or single hand-crafted feature based model alone (DFFN \emph{w/o}  CNN). Hence, the fusion of deep features and hand-crafted features helps in boosting the performance.
\begin{table*}[!t]
\centering
\begin{tabular}{p{4.25cm}@{\hskip 0.25cm} p{3cm}@{\hskip 0.2cm} p{0.8cm}@{\hskip 0.05cm} p{0.6cm}@{\hskip 0.35cm} p{0.65cm}@{\hskip 0.35cm} p{0.65cm}@{\hskip 0.35cm} p{0.65cm}@{\hskip 0.2cm} p{5.5cm} }
\hline
\textbf{ Question } & \textbf{Answer} &  \fontsize{7.3}{8} \selectfont\textbf{True Label} &  \fontsize{7}{8} \selectfont\textbf{DFFN} & \fontsize{7}{8} \selectfont\textbf{JAIST} & \fontsize{6.5}{8} \selectfont\textbf{HITSZ ICRC} & \fontsize{6.5}{8} \selectfont\textbf{ICRC HIT} & \textbf{Comment}\\ 
\hline

Can anyone plz help me this problem? I need to send a mobile phone to (Jaipur) India. I  contacted DHL but they are charging very high. Is there any other company like DHL? Plz specify... & You can send by post office for cheap price (compare to Courier service) & Good & Good & Bad & Pote-ntial & Pote-ntial &  \textbf{\textit{GCD}} similarity  feature captures that \textit{post office, DHL, courier} are linked to similar pages when they occur as anchor texts.\\
\hline
I have a valid work permit not expiring until next  year. My question is will a passport with less than 6 months be a problem when entering the country again? has anyone got an idea or experienced something similar to this situation? & You can ask your embassy if they can give you temporary extension? In our embassy, we have an option to extend & Good & Good & Bad & Pote-ntial & Good & \textbf{\textit{Sent2Vec}} has good similarity score, as \textit{embassy, passport} co-occur in search queries. CNN based features also contributed to the score.\\
\hline
What softwares are you using for downloading movies? I'm using limewire and utorrent. How about you? & im using azureus client..limewire sucks (lol) & Good & Good & Bad & Good & Bad & \textbf{\textit{TagMe}} links \textit{azureus, limewire, utorrent} to their wiki pages and finds out that they all belong to movie torrent software class. \\
\hline 
I want to take my car to India for vacation on Qatar license plate and drive there a couple of months and bring it back again to Qatar. & Have seen a couple of Qatar REGN. vehicles in Kerala recently.  & Bad & Good & Good &  Good & Good & \textbf{\textit{Sent2Vec}} has high similarity scores as \textit{regn vehicles, license plate, car} co-occur in search queries. \\
\hline
I saw a little girl running by the streets , and she had a cat attached to her ......is that normal in this country? & I saw a little girl running by the streets , and she had a parent attached to her ......is that normal in this country?  & Bad & Pote-ntial & Pote-ntial & Bad & Bad & \textbf{\textit{Author Reputation}}  gave neutral sim. score as the answer author had written very less answers in the forum and had almost equal number of Good and Bad answers.  \textbf{\textit{Question Authors' Response Pattern}} also gave a neutral score as question author has commented before and after this answer. \textbf{\textit{Wiki based}} and \textbf{\textit{Sent. Vec.}} features gave high scores as question and answer are exactly same except for one word. \\

\hline
\end{tabular}
\caption{Qualitative Analysis of DFFN Results with respect to other baseline approaches.}
\label{qualitative}
\end{table*}

\subsection{Qualitative Analysis}
In Table \ref{qualitative}, we present a qualitative analysis of DFFN results comparing it other baselines. The first three examples are cases where we correctly predicted the target label. In the last two examples, we incorrectly predict the label.

In the first example, all the baseline systems predict incorrectly while DFFN is able to predict correctly as Good as the \textit{GCD} similarity feature captures that \textit{post office, DHL, courier} are linked to similar pages when they occur as anchor texts. In the second example, JAIST and HITSZ-ICRC output incorrectly while DFFN does it correctly as \textit{Sent2Vec} has good similarity score, as \textit{embassy, passport} co-occur in search queries. CNN based features also contributed to the score. In the third example, JAIST and ICRC-HIT get it incorrectly whereas DFFN is able to predict correctly as \textit{TagMe} links \textit{azureus, limewire, utorrent}, to their wiki pages due to which it can find out that they all belong to movie torrent software class. 

In the fourth example, all the systems predict it wrongly as Good instead of Bad. In the case of DFFN, this is due to \textit{Sent2Vec} scoring a high similarity score as \textit{regn. vehicles, license plate, car} co-occur in search queries. In the last example HITSZ-ICRC and ICRC-HIT predict it correctly while DFFN predicts it wrongly as Potential instead of Bad. For this example \textit{Author Reputation} feature gave a neutral similarity score as the answer author had written very less answers in the forum and had almost equal number of Good and Bad answers (Good:4, Potential:1, Bad:5). \textit{Question Authors' Response Pattern} feature has also given a neutral similarity score as question author commented before and after this answer. \textit{Wikipedia based} and \textit{Sentence Vector} features have given high similarity scores since question and answer are exactly the same except for one word as this was answered in a sarcastic way. Thus DFFN predicted Potential in this example.

\section{Conclusion}\label{conclusion}
We present a novel approach \emph{``Deep Feature Fusion Networks (DFFN)''} which combines HCF features into a CNN model for improving answer quality prediction. DFFN enriches the feature representations learnt through a CNN by introducing more similarity features computed using external resources such as Wikipedia, Google Cross-Lingual Dictionary (GCD), Clickthrough Data. As a result, we show that DFFN achieves state-of-the-art performance on the standard SemEval-2015 and SemEval-2016 benchmark datasets and shows better performance than baseline approaches which individually employ either HCF or DL based techniques. In future work, we would like to investigate the difference between features learnt using DFFN and a stand alone CNN.
\bibliographystyle{plain}
\bibliography{citations.bib}
\end{document}